**Modeling granular material segregation using a combined finite element method and advection-diffusion-segregation equation model**

Yu Liu[a], Marcial Gonzalez[a,c] and Carl Wassgren[a,b,*]

[a]School of Mechanical Engineering, 585 Purdue Mall, Purdue University, West Lafayette, IN 47907-2088, U.S.A.
[b]Department of Industrial and Physical Pharmacy (by courtesy), 575 Stadium Mall Drive, Purdue University, West Lafayette, IN 47907-2091, U.S.A.
[c]Ray W. Herrick Laboratories, Purdue University, West Lafayette, IN 47907, U.S.A.
* Corresponding author at: School of Mechanical Engineering, Purdue University, West Lafayette, IN 47907-2088, U.S.A., Tel.: +1 765 494 5656, *E-mail address*: wassgren@purdue.edu (C. Wassgren).

**Abstract**
A two-dimensional, transient, multi-scale modeling approach is presented for predicting the magnitude and rate of percolation segregation for binary mixtures of granular material in a rotating drum and conical hopper. The model utilizes finite element method simulations to determine the bulk-level granular velocity field, which is then combined with particle-level diffusion and segregation correlations using the advection-diffusion-segregation equation. The utility of this modelling approach is demonstrated by predicting segregation patterns in a rotating drum and during the discharge of conical hoppers with different geometries. The model exhibits good quantitative accuracy in predicting DEM and experimental segregation data reported in the literature for cohesionless granular materials. Moreover, since the numerical approach does not directly model individual particles, it is expected to scale well to systems of industrial scale.

## 1 Introduction and background

Granular materials are processed in many industries, such as those that manufacture chemicals, food products, and pharmaceuticals. Unintentionally heterogeneous powder blends can result in inconsistencies during processing and unacceptable product quality. The components of a granular mixture typically have different properties, such as size and shape, which can result in the segregation of the components. Hence, it is useful to have tools for predicting segregation in order to help better design and manage unit operations and, ultimately, product quality.

Many phenomena can result in segregation, such as vibration-driven percolation and convection [1–3] and elutriation [4]. In particular, gravity-driven segregation in dense, sheared granular flows, referred to here as shear-driven percolation [5–7], is a common mechanism that occurs during industrial processing. In shear-driven percolation, compared to large particles, small particles have an increased probability of falling through gaps that form between particles when the particle assembly is subject to shear. As a result, smaller particles collect below the shear layer leaving the larger particles at the top [5,6,8–12].

A number of studies have modeled shear-driven percolation segregation using a continuum approach that incorporates advection due to mean flow, percolation-driven segregation, and diffusion [5,13–18]. Most of these models were used to gain a qualitative understanding of the complex physics while some showed good agreement with experiments [19,20]. Recently developed continuum models utilized discrete element method (DEM) simulations to derive particle diffusion and segregation correlations at a local, i.e., particle-level, scale and combined these correlations with analytically-derived advection fields at the macro-scale [21–23]. Predictions from these studies were shown to be quantitatively accurate when compared with DEM-only simulations and experiments.

A variety of flows have been studied using a continuum approach to obtain advection fields, such as chute [17,19,21], plug [14], annular shear [16], and rotating drum [22] flows. However, the domains of these flows were simple, two-dimensional, steady geometries amenable to analytical solutions for the macroscopic flow field. Indeed, the shear layers in these geometries are often approximated as having linear or exponential velocity profiles located at a free surface. To study more complex geometries, a computational approach is needed for obtaining the macroscopic flow field. For continuum modeling, this means that a constitutive model describing the stress-strain-strain rate behavior is required. Constitutive models have been developed to describe granular flow dynamics, such as the Schaeffer model [24,25], the μ(I) model [26–28], and the hydrodynamic model [29–31]. These models have been used to predict the flow behavior of granular materials in more complex configurations than those studied analytically, such as silos and hoppers [27,29,30], a high-shear granulator [28], and an asymmetric double cone mixer [31]. Although good agreement with DEM simulations and experiments have been observed for several aspects of these flows, such as velocity fields and wall stress profiles [30,32], these flows were still mainly restricted to two-dimensional geometries. Recently, finite element method (FEM) simulations with Mohr-Coulomb [33–35] and Drucker–Prager [36–38] constitutive material models have been used to study the granular flow behavior in both two- and three-dimensional systems, and quantitatively accurate predictions were observed. The advantages of using FEM simulations with an elasto-plastic material model over previous constitutive models are that (a) unsteady granular flows in three-dimensional configurations can be simulated, and (b) experimental characterization of the required material properties is usually straightforward using, for example, standard shear cell equipment.

Only recently have researchers begun to combine computationally predicted velocity fields with expressions for particle diffusion and segregation. For example, Bertuola et al. [39] predicted segregation in a discharging two-dimensional hopper using segregation correlations derived by Fan et al. [21] and Hajra et al. [40] combined with flow field predictions using a hydrodynamic model for particle flow. The model was able to quantitatively predict the degree of segregation compared with published experiments after key model parameters were fitted to the experimental data. It is worth noting that the hydrodynamic model used to simulate the macroscopic flow behavior was less accurate than the one predicted by a Mohr-Coulomb model [37,38]. Bai et al. [41] used an FEM model with Mohr-Coulomb constitutive behavior to predict the degree of blending in a cylindrical, bladed mixer assuming convective mixing only. The result was observed to be mesh-size dependent. Liu and co-workers [42,43] recently developed a multi-scale model that combines particle diffusion coefficient correlations with advective flow field information from FEM simulations using a Mohr-Coulomb constitutive model. The model was able to quantitatively predict the magnitude and rate of powder blending in a steady, two-dimensional rotating drum and an unsteady, but periodic, three-dimensional Tote blender. No backfitting of parameters was required. Segregation was not included in the model since all of the particles had identical properties.

This work extends the model developed by Liu et al. [42,43] to include segregation. FEM simulations with a Mohr-Coulomb elasto-plastic material model are used to provide a prediction of the advective flow field. This flow field is combined with particle-level diffusion and shear-driven percolation segregation correlations to predict segregation in a rotating drum and hoppers of different geometries. The predictions are compared to DEM [22] and experimental [44] results available in the literature. Section 2 introduces the FEM modeling approach and its numerical implementation. Section 3 describes the advection-diffusion-segregation equation used in the model. Section 4 presents comparisons of the model predictions to the DEM and experimental results.

## 2      Finite element method model

Three-dimensional, coupled Eulerian-Lagrangian, FEM models [42,43] are used here to predict the advective flow field in a rotating drum and conical hoppers. Previous work [36–38,42,43,45–47] has shown that FEM models can accurately predict granular material behavior advective flow fields [36–38,42,43]. The following sub-sections describe the model geometries, boundary conditions, and initial conditions for three different systems.

### 2.1     Simulation of a rotating drum

The commercial FEM package Abaqus/Explicit V6.14 is used to perform the simulations. The geometry of the simulated rotating drum, shown in Figure 1, is based on previous DEM simulations performed by Schlick et al. [22] and it mimics a lab-scale rotating drum with a diameter of 150 mm. Since a 2-D flow pattern was assumed in [22], both the front and back sides of the Eulerian mesh are regarded as planes of symmetry in the model with a narrow width of 10 mm used for computational efficiency. Gravity is included in the model with $g$ = 9.8 m/s$^2$ directed in the negative $y$-direction. The rotational speed is 0.75 rad/s (7.2 rpm), corresponding to the previously published work [22].

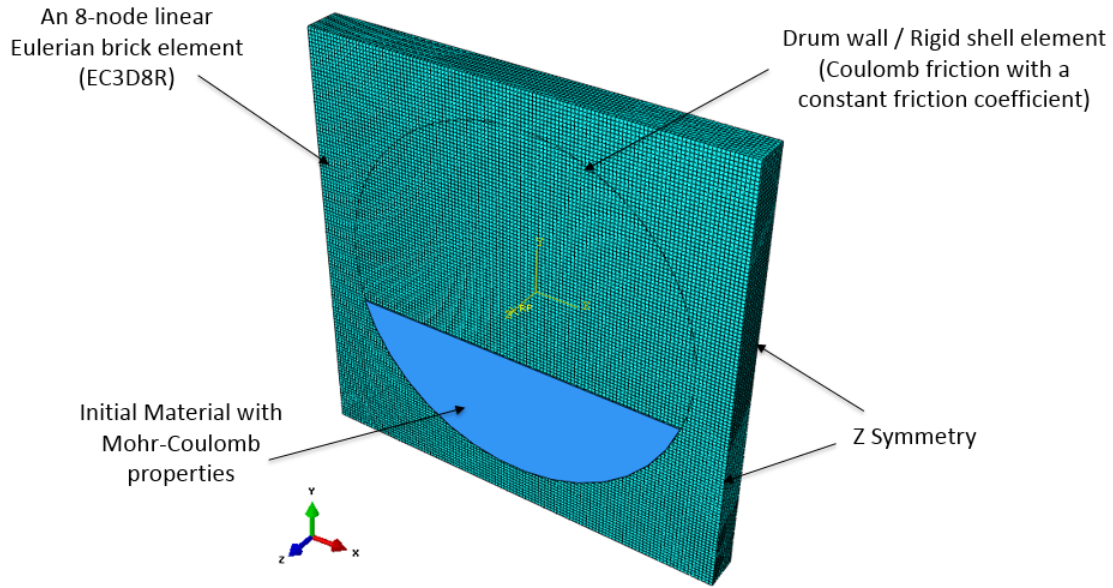

Figure 1. A schematic of the geometry modeled in the FEM simulations.

A Mohr-Coulomb elasto-plastic (MCEP) model is used in the current work to describe the stress-strain behavior of the particulate material. Previous research has shown that the MCEP model can accurately describe the behavior of dense, flowing granular materials [36–38,42,43]. Note that this constitutive model is shear rate independent and the current implementation does not take into account changes in material porosity. Also, the MCEP model cannot predict the formation of shear bands without considering shear localization. Other constitutive models can be used when such factors are required to improve the model's accuracy [48]. Despite the simplicity of the MCEP model, it is shown in the Results section to be sufficiently accurate at predicting velocity fields used in the quantitative prediction of segregation trends.

The material properties needed in the MCEP model are bulk density, Young's modulus, Poisson's ratio, dilation angle, material internal friction angle, and wall friction angle. All of these material parameters can be obtained from independent, standard material tests. For example, a uniaxial compression test can be used to calibrate the bulk density, Young's modulus, and Poisson's ratio, while a shear cell test can be used to calibrate the material's internal friction angle and wall friction angle. Refer to Liu et al. [42], and references therein, for a detailed description of these experimental techniques. As a side note, the effort required to obtain model parameters should not be underestimated. More complex material models can require many parameters, some of which may be difficult to obtain. For example, in the $\mu(I)$ model [28], the friction coefficient must be measured as a function of the inertial number, and the switching inertial number, which sets the transition from dilute to dense granular flow, must be calibrated. Determining these parameters is not trivial. In addition, many of the more complex material models are not implemented in commercial software, making their use for industrial practioners challenging. The MCEP provides a good balance between model accuracy, simplicity, and is already implemented in commercially available software and, thus, is worth consideration.

The Abaqus element mesh for the rotating drum is shown in Figure 1 and derived from the model described in [42]. A Coupled Eulerian-Lagrangian (CEL) approach is used to handle the interactions between Eulerian and Lagrangian elements. The Eulerian Volume Fraction (EVF) value is used to determine the volume of material within each element. A value of EVF = 0 indicates that no material is present in the element while EVF = 1 indicates that the element is completely filled with material. EC3D8R (8-node linear hexahedron) elements are used in the current work with a reduced integration scheme to prevent locking [49]. Gravity is increased gradually to fill the drum and allow material to settle before the drum rotates. Further details on the model set up can be found in previously published work [42].

**2.2 Simulations of conical hoppers**

Figure 2 shows the geometries of the simulated conical hoppers, which correspond to the hoppers used by Ketterhagen et al. [44] in their experimental work. The FEM discretization of these three-dimensional geometries is shown in Figure 3. A symmetry boundary condition is applied on the front side of the Eulerian mesh, as shown in Figure 3, and only half of the geometry is modelled to save computational time. Note that an axisymmetric boundary condition cannot be applied in the current model since it causes numerical errors along the axisymmetric axis (refer to Section 3.2). Hopper walls are modeled as rigid shells and are fixed in all degrees of freedom. Gravity is included in the model with $g = 9.8$ m/s$^2$ directed in the negative $y$-direction.

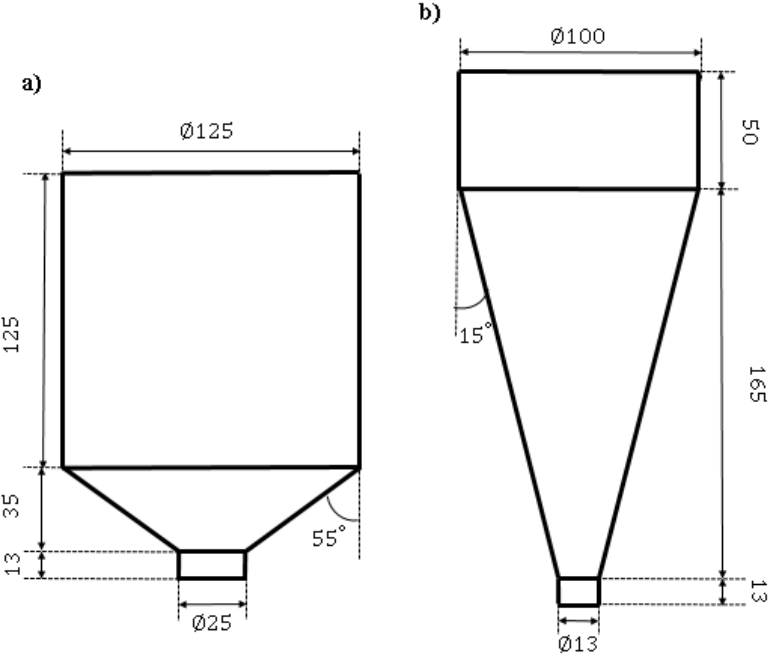

Figure 2. Schematics and dimensions of the experimental hoppers used by Ketterhagen et al. [44]. Length dimensions are in mm.

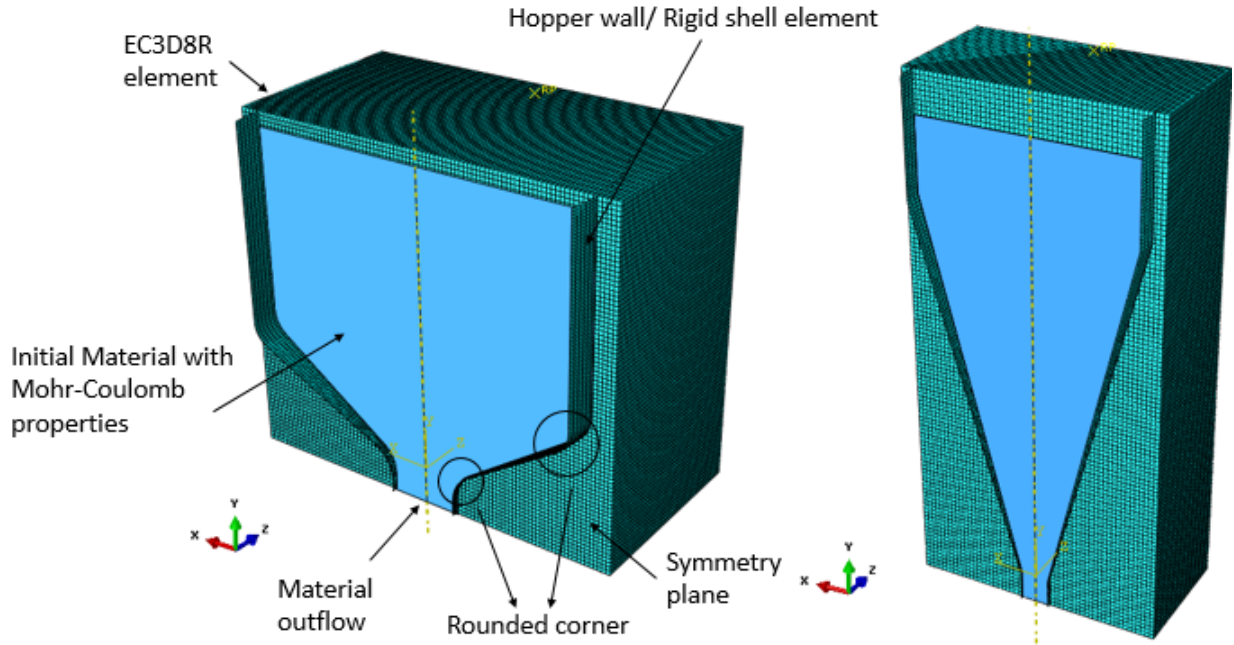

Figure 3. The discretization of the computational hopper domains.

Details of the material model and Abaqus implementations are the same as those described in Section 2.1. The hopper outlet is initially closed and, after the material settles under gravity within the Eulerian elements, the outlet is opened and discharge commences. Specifically, the outlet is opened by assigning a free-flow Eulerian boundary condition to the bottom plane, as shown in Figure 3.

It is worth noting that since a coupled Eulerian-Lagrangian (CEL) method is used, the Lagrangian mesh, i.e., the hopper wall, is placed inside the Eulerian mesh. A penalty method is then used to prevent material penetration and, thus, to ensure mass conservation. The algorithm calculates a repulsive contact force proportional to the penetration distance between the Lagrangian mesh and the material in the Eulerian mesh [50]. Naturally, a penalty method cannot strictly enforce the constraint and, hence, some penetration of Eulerian material into the Lagrangian boundary occurs. In most cases, this penetration is negligible; however, depending on the system geometry and material properties, severe penetration can occur in the simulation. There are a number of modeling procedures to overcome these severe cases, namely: (1) use a refined mesh in the region where penetration happens; (2) reduce the time step size so that a smaller penetration distance is used to calculate the contact force; and (3) round sharp corners of the Lagrangian mesh, as shown in Figure 3.

## 3    The multi-scale segregation model

Since a quasi-2D rotating drum flow [22] and axisymmetric conical hopper flows [44] are studied in this work, a two-dimensional segregation model is developed as an extension of the two-dimensional blending model in [42]. The following sub-sections present the main aspects of the segregation model.

## 3.1 Advection-diffusion-segregation (ADS) equation

The advection-diffusion-segregation (ADS) equation is used to model the diffusion and shear-induced percolation segregation of a binary granular mixture and the resulting temporally and spatially varying concentration fields of the component materials. Specifically, the governing equation is,

$$\frac{\partial c_i}{\partial t} = \nabla \cdot (\boldsymbol{D} \nabla c_i) - \nabla \cdot (\boldsymbol{v} c_i) - \nabla \cdot (\boldsymbol{v}_p c_i), \tag{1}$$

where $c_i$ is the local concentration of material species $i$ (either small, $i = s$, or large, $i = l$, particles). The parameter $\boldsymbol{D}$ is the diffusion coefficient tensor for that species, $\boldsymbol{v}$ is the local advective velocity vector of the bulk material, and $\boldsymbol{v}_p$ is the percolation velocity vector. Since binary mixture is studied in the current work, $\boldsymbol{D}$ and $\boldsymbol{v}_p$ represent the mixing and segregation parameters between the two species. Note that previous work has shown that the self-diffusion coefficient $\boldsymbol{D}$ is an anisotropic tensor quantity [42,43,51]. However, Fan et al. [21] showed that for segregation-dominated flows, a constant $D$, namely, the mean diffusion coefficient in the spanwise direction, can still lead to an accurate prediction. Hence, for simplicity, a constant diffusion coefficient $D$ is used in the current model, which is assumed independent of particle size, shear rate, and local concentration, consistent with previous work [21].

The percolation velocity $\boldsymbol{v}_p$ derived by Fan et al. [21] is adopted in the current model. In their work, heap flows were studied and, thus, only the normal component of the percolation velocity relative to the mean normal flow was considered. The streamwise component was neglected. Here, gravity acts in the negative $y$-direction and percolation is dominant in the direction of gravity [39,52]. Therefore, the $x$-component of the percolation velocity is neglected and only the $y$-component is considered. Moreover, according to Fan et al. [21], the percolation velocity can be approximated as a linear function of the shear rate and the concentration of the other species in a bi-disperse mixture, i.e.,

$$v_{p,s} = -S|\dot{\gamma}|(1-c_s), \quad v_{p,l} = S|\dot{\gamma}|(1-c_l), \tag{2}$$

where $S$ is the percolation length scale and $|\dot{\gamma}|$ is the magnitude of the spanwise shear rate. Note that unlike the diffusion coefficient, the percolation speed does depend on the local particle concentration.

Relationships for the percolation length scale $S$ as a function of the particle diameter and small to large particle ratio have been proposed [21,22,39,40]. However, in this work, a percolation length either previously reported [22] or fitted to experimental data [44] is used. Finally, since the percolation in the $y$-direction is mainly caused by the shear rate in the $x$-direction, the $y$-component of the shear rate is neglected and the shear rate is approximated by,

$$|\dot{\gamma}| \approx |\dot{\gamma}_x| = \left|\frac{\partial v_x}{\partial y}\right|. \tag{3}$$

Using the local mass conservation equation, adopting the relationship presented above (Eqs. (2) and (3)), and assuming an incompressible material, i.e.,

$$\nabla \cdot \boldsymbol{v} = 0, \tag{4}$$

Eq. (1) may be written in index notation form as,

$$\frac{\partial c_i}{\partial t} = D\left(\frac{\partial^2 c_i}{\partial x^2} + \frac{\partial^2 c_i}{\partial y^2}\right) - \left(v_x \frac{\partial c_i}{\partial x} + v_y \frac{\partial c_i}{\partial y}\right) - \frac{\partial\left[\pm S \left|\frac{\partial v_x}{\partial y}\right|(1-c_i)c_i\right]}{\partial y}, \quad (5)$$

where the $\pm$ sign is determined by the size of the particles, as indicated in Eq. (2).

## 3.2 Numerical method

The numerical method used to solve the ADS equation in the current model is the same one used in previous works [42,43]. A finite difference method based on a second-order Tylor Lax-Wendroff scheme is used to solve Eq. (5) due to the method's simplicity and computational efficiency [53]. The second order scheme can be written as,

$$\begin{aligned}c_{ij}^{n+1} = c_{ij}^n &- \left[v_x \Delta_{x0} c_{ij}^n - \left(\frac{1}{2}v_x^2 + \mu_x\right)\delta_x^2 c_{ij}^n\right] - \left[v_y \Delta_{y0} c_{ij}^n - \left(\frac{1}{2}v_y^2 + \mu_y\right)\delta_y^2 c_{ij}^n\right] \\ &+ v_x v_y \Delta_{x0}\Delta_{y0} c_{ij}^n \mp S\left[\Delta_y \left|\Delta_y (v_x)_{ij}\right|(1-c_{ij}^n)c_{ij}^n + \left|\Delta_y (v_x)_{ij}\right|(1-2c_{ij}^n)\Delta_y c_{ij}^n\right]\Delta t\end{aligned} \quad (6)$$

where,

$$v_x = \frac{\left[(v_x)_{ij} + \Delta_{x0}(D_{xx})_{ij}\right]\Delta t}{\Delta x}, \quad (7)$$

$$v_y = \frac{\left[(v_x)_{ij} + \Delta_{y0}(D_{yy})_{ij}\right]\Delta t}{\Delta y}, \quad (8)$$

$$\mu_x = (D_{xx})_{ij}\frac{\Delta t}{\Delta x^2}, \quad (9)$$

$$\mu_y = (D_{yy})_{ij}\frac{\Delta t}{\Delta y^2}, \quad (10)$$

$$\Delta_{x0} c_{ij}^n = \frac{c_{i,j+1}^n - c_{i,j-1}^n}{2}, \quad (11)$$

$$\Delta_{y0} c_{ij}^n = \frac{c_{i+1,j}^n - c_{i-1,j}^n}{2}, \quad (12)$$

$$\delta_x^2 c_{ij}^n = c_{i,j+1}^n - 2c_{ij}^n + c_{i,j-1}^n, \quad (13)$$

$$\delta_y^2 c_{ij}^n = c_{i+1,j}^n - 2c_{ij}^n + c_{i-1,j}^n, \quad (14)$$

$$\Delta_{x0}\Delta_{y0} c_{ij}^n = \frac{\left(c_{i+1,j+1}^n - c_{i-1,j+1}^n - c_{i+1,j-1}^n + c_{i-1,j-1}^n\right)}{4}, \quad (15)$$

$$\Delta_{x0}(D_{xx})_{ij} = \frac{(D_{xx})_{i,j+1} - (D_{xx})_{i,j-1}}{2}, \tag{16}$$

$$\Delta_{y0}(D_{yy})_{ij} = \frac{(D_{yy})_{i+1,j} - (D_{yy})_{i-1,j}}{2}, \tag{17}$$

$$\left|\Delta_y(v_x)_{ij}\right| = \left|\frac{(v_x)_{i+1,j} - (v_x)_{i-1,j}}{2 \cdot \Delta y}\right|, \tag{18}$$

$$\Delta_y\left|\Delta_y(v_x)_{ij}\right| = \frac{\left|\Delta_y(v_x)_{i+1,j}\right| - \left|\Delta_y(v_x)_{i-1,j}\right|}{2 \cdot \Delta y}, \text{ and} \tag{19}$$

$$\Delta_y c_{ij}^n = \frac{c_{i+1,j}^n - c_{i-1,j}^n}{2 \cdot \Delta y}. \tag{20}$$

The computational molecule for this scheme is shown in Figure 4 of Liu et al. [42]. Note that since a finite difference method is used, the Von Neumann conditions must be checked to ensure the stability of the numerical computations.

A MATLAB program is used to iterate the finite difference scheme. A C++ code was developed to read and process the large FEM output files (.obd files). The material boundaries computed within the FEM simulation from the Eulerian Volume Fraction (EVF) values are used directly by the MATLAB algorithm. To ensure mass conservation of each species, the segregation and diffusion fluxes are set to zero on the boundary and, thus, material cannot exit the domain by advection. A threshold is set to ensure the material concentration value remains between zero and one, and a small time step is carefully chosen to ensure the stability of the explicit scheme. Finally, a fitting approach is applied to ensure that the material concentration of the boundary node is equal to the value of the node that is one grid point inward, a scheme commonly employed in CFD computations [54]. Note that this fitting approach restricts the use of an axisymmetric boundary condition since the material nodes along the axisymmetric axis would be treated as boundary nodes as well.

Mesh dependency studies were performed to ensure the convergence of the numerical results for both the FEM and the ADS equation calculations. For the FEM simulations, comparisons of the averaged velocity differences in the system were compared to determine convergence, with the details shown in Table 1. The velocity differences were averaged between 10 different points along the free surface of the rotating drum and the outlet of the hopper. For both systems, the FEM solutions were insensitive to the mesh sizes and, thus, 500,000 and 176,000 elements were used in the rotating drum simulation and hopper simulations, respectively. Detailed mesh convergence results for the ADS equation calculations are given in Sections 4.1 and 4.2.

Table 1. Convergence study results for the FEM simulations.

|  | Number of elements | | Averaged velocity differences |
|---|---|---|---|
|  | Case 1 | Case 2 |  |
| Rotating drum | 500,000 | 1,280,000 | 2.74% |
| 15° hopper | 176,000 | 411,000 | 3.27% |

Note that to achieve numerical convergence and stability, the number of nodes used in the second-order Tylor Lax-Wendroff scheme must be much larger than the number of nodes in the FEM mesh. A linear interpolation algorithm, implemented in MATLAB, is used to transfer data from the FEM nodes to the ADS nodes.

The initial conditions used in the simulations correspond to a perfect mixture since segregation is the main focus of this work. The reader is referred to [42,43] for examples of granular systems that are initially partially mixed or separated and are then blended.

## 4      Results

In this section, predictions from the FEM-ADS equation models are compared to previously published DEM and experimental results. Specifically, the segregation profile normal to the bed surface is examined for a rotating drum and the temporal variation in the fraction of fine particles at discharge is compared for two different hoppers. In addition, qualitative examination of the particle concentration fields are discussed as well as parametric studies.

### 4.1    Rotating drum

The rotating drum FEM-ADS model predictions are compared to bi-disperse particle segregation DEM results reported by Schlick et al. [22]. The parameters used in the FEM simulations are given in Table 2 and correspond to the properties derived from DEM simulations of 1 mm diameter, identical spherical particles with the material properties given in Liu et al. [42]. The FEM-ADS models described in this paper are one-way coupled, which means that the bulk flow field determined from the FEM model is unaffected by the local particle species concentration.

There are two differences between these previous DEM simulation parameters and those of the Schlick et al. work [22]. First, Schlick et al. used a particle-wall friction coefficient of 0.4 while the Liu et al. [42] work used a value of 0.3. This difference is expected to have little impact since Liu et al. [42] demonstrated that changing the particle-wall friction coefficient had little impact on the flow behavior. The reason is that the first avalanche always occurs at the same location and the free surface angle remains constant as long as the wall friction is sufficiently large to lift the material without sliding. Second, the Schlick et al. work used a 50/50 bi-disperse assembly of 1 mm and 3 mm spheres while the DEM simulations used by Liu et al. [42] to calibrate the FEM parameters used identical 1 mm particles.

The dilation of cohesionless granular materials is usually small and, thus, a dilation angle of 0.1° was adopted, which is the minimum value allowed in Abaqus for the MCEP model. Note that even with a small dilation angle, the accumulated bed dilation will grow with large shear strains. However, for the current work, the dilation is not significant since the FEM velocity field is periodic for the rotating drum and so the accumulated shear strain is small. Similarly, for the hopper flow the total shear train is also small so that bed dilation is negligible. Simulations that involve large shear strains should consider the use of a different constitutive model that does not result in excessive bed dilation. The Poisson's ratio is very small because the powder bed is loose, with a relative density around 0.3, and thus it is very compressible. Previous work has demonstrated that small Poisson's ratios are measured at small solid fractions [45,47,55].

Table 2. Parameters used in the rotating drum FEM simulation.

| Parameter | Value |
| --- | --- |
| Material density (kg/m$^3$) | 1500 |
| Young's modulus (MPa) | 3.65 |
| Poisson's ratio (-) | 0.065 |
| Internal friction angle (degree) | 23.6 |
| Cohesion (Pa) | 0 |
| Dilation angle (degree) | 0.1 |
| Wall friction coefficient (-) | 0.324 |

Schlick et al. used particles with 1 mm and 3 mm diameters in their DEM simulations [22]. As mentioned previously, the FEM material parameters were calibrated using 1 mm particles only. To verify that the FEM model predicted the flow field accurately, the surface velocity in the streamwise direction of the DEM simulation reported by Schlick et al. was compared with that of an FEM simulation using the material parameters in Table 2 when the rotation speed $\omega$ is 0.75 rad/s (7.2 RPM). Figure 4 shows that these two velocities are similar despite having different particle sizes. The maximum thickness of the flowing layer $\delta_0$ was also compared and the values predicted by the DEM and FEM simulations differ by, at most, 8% (14.8 mm for the former and 13.6 mm for the latter). This close similarity indicates that the Mohr-Coulomb properties listed in Table 2 represent the granular system used in the DEM simulations with sufficient accuracy for the rotating drum studied. This similarity may not hold true for other geometries with large particle size differences since previous work has shown that the velocity field can be influenced by the particle size [22].

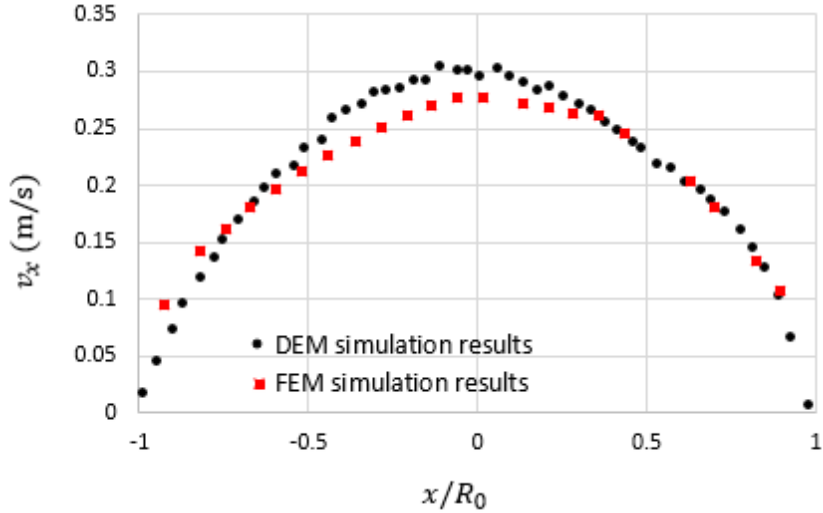

Figure 4. Surface velocity as a function of the streamwise position at $\omega = 0.75$ rad/s (7.2 RPM). The coordinate system used in the plot is identical to the one used by Schlick et al. [22] for consistency, where $x$ is the streamwise direction in the flowing layer.

As mentioned in Section 3, it is necessary to know the diffusion coefficient $D$ and the percolation length scale $S$ in order to compare results from the FEM-ADS model with those from DEM simulations. These values were found in [22] for rotating drum DEM simulations of 1 mm and 3 mm diameter spheres with $\omega = 0.75$ rad/s (7.2 RPM) to be $D = 16.1$ mm$^2$/s and $S = 0.29$ mm.

Figure 5 shows the spatial evolution of the small particle concentration at different times predicted by the FEM-ADS model. The drum is half-filled with initially well-mixed particles, i.e., $c_s = c_l = 0.5$ at every material point in the domain. As expected for a bi-disperse granular system, as time increases the degree of segregation increases. It is evident from the figure that small particles segregate to the bottom of the flowing layer and gather in the center of the material domain inside the drum. The same trend was also captured in the previous DEM simulations of Schlick et al. [22]. Hence, qualitatively, the multi-scale model reproduces the segregation pattern observed in the DEM simulations.

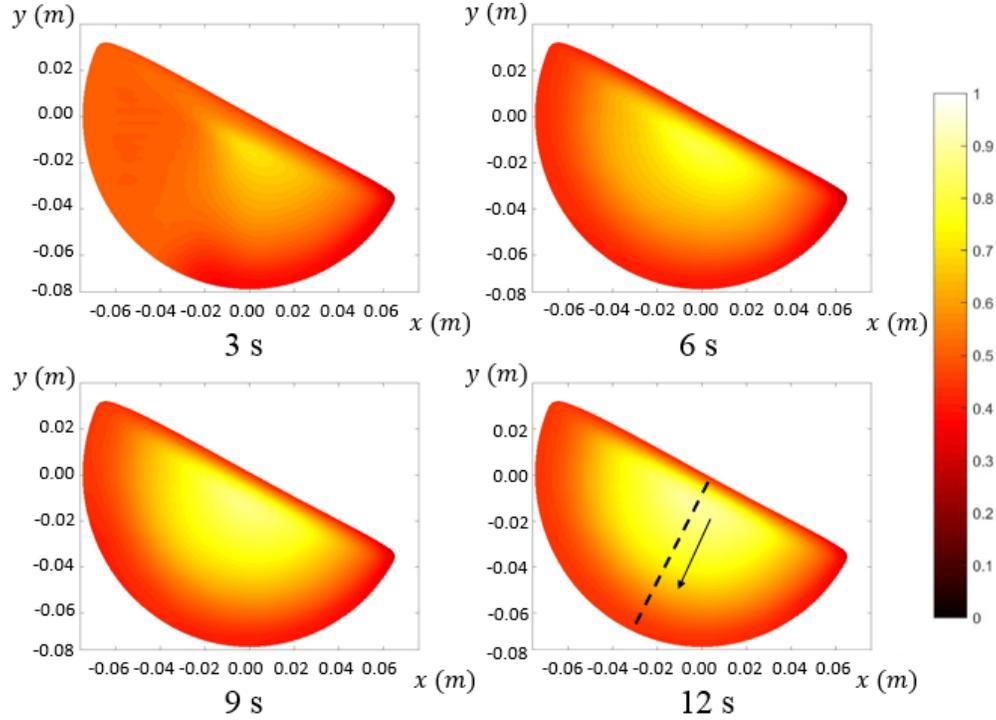

Figure 5. Snapshots showing the small particle concentration in the simulated rotating drum at different times. The dashed line at 12 s is the path used to plot the small particle concentration in Figure 6.

To provide a quantitative comparison of the two modelling approaches, the steady-state concentration of small particles is plotted in Figure 6 as function of the dimensionless distance $\delta / R_0$ along the center of the bed starting from the free surface, i.e., along the dashed line shown in Figure 5. The distance is made dimensionless using the radius of the drum $R_0$. The figure indicates that there is good quantitative agreement between the two models, although the FEM-ADS model slightly overpredicts the small particle concentration near the drum walls. The total wall-clock time required to run the FEM (16 cores with the Intel Xeon CPU E5-2680 v3 processor) and ADS (MATLAB, single core with the same processor) simulations was approximately two days. Note that a mesh dependency study was performed to ensure the convergence of the ADS equation calculations. The small particle concentration along the center of the bed, as shown in Figure 6, was computed using 250,000 and 1,000,000 elements, respectively. The averaged error among all computed data points was 3.56% and, hence, 250,000 elements were used in the remainder of the rotating drum simulations.

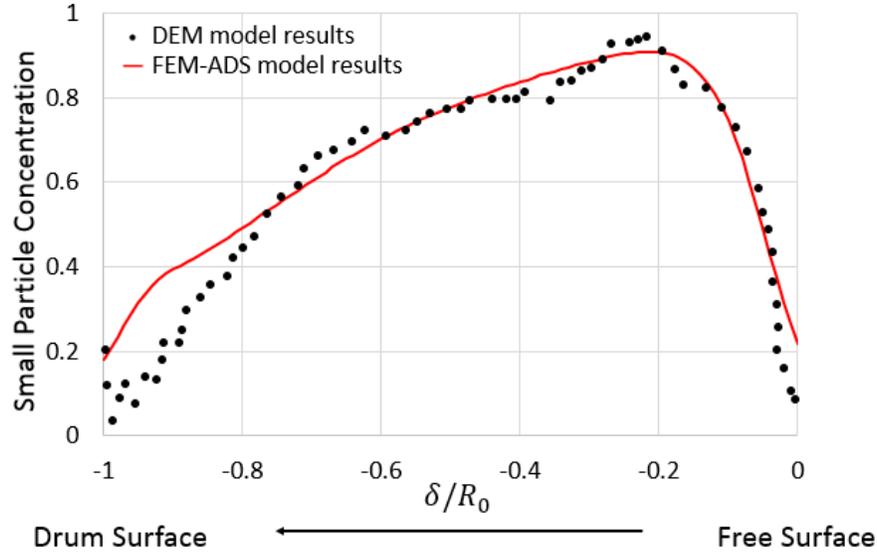

Figure 6. Steady-state concentration of small particles as a function of dimensionless distance from the free surface, $\delta/R_0$, along the centerline of the drum. The DEM model results are from previous work by Schlick et al. [22].

To better understand the effects of the model parameters, a parametric study was performed using the rotating drum simulation. The model is identical to the one described previously except for the values of the diffusion coefficient $D$ and percolation length scale $S$. Figure 7(a) shows the results for simulations where the percolation length scale remains the same while the diffusion coefficient changes and Figure 7(b) shows the results when the diffusion coefficient remains the same and the percolation length scale changes. Clearly, segregation is stronger as the diffusion coefficient decreases and percolation length increases. This same trend was predicted by Schlick et al. [22]. It is also shown that segregation is dominated more by the percolation than the diffusion, and the effect of percolation and diffusion decreases as the powder bed approaches a fully segregated state. Moreover, it is noticed in both Figure 7(a) and (b) that the maximum small particle concentration occurs at the bottom of the active layer since small particles fall to the bottom of the flowing layer as they move downstream and gather in the center of the bed. Also, when the diffusion coefficient is large or the percolation length scale is small, there is a small "mixing band" at around $\delta/R_0 = -0.9$. This behavior occurs as a result of the small velocity gradient caused by wall friction close to the drum wall. A large diffusion coefficient and small percolation length scale results in more mixing in this region.

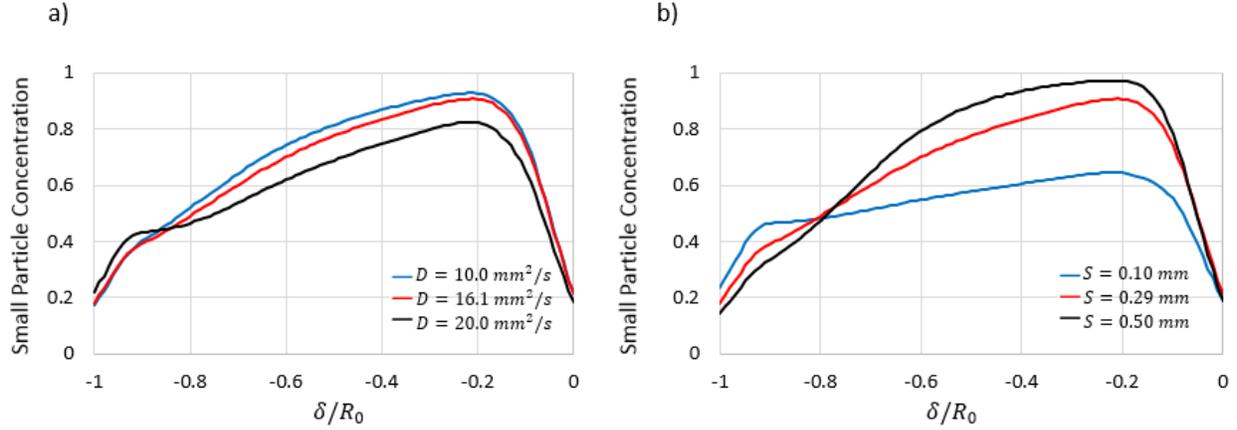

Figure 7. The steady state small particle concentration plotted as a function of dimensionless distance from the free surface, $\delta/R_0$, along the centerline of the drum for a) different diffusion coefficients $D$ and b) different percolation lengths $S$. Other model parameters are the same as those used in Table 2.

## 4.2 Conical hoppers

Experimental work on bi-disperse particle segregation carried out by Ketterhagen et al. [44] was used to further validate the predictions of the FEM-ADS model. The experimental setup consists of bench scale hoppers (ASTM D 6940-03) and binary mixtures of glass beads, as shown in Figure 2. The initial hopper fill height is 105 mm for the 55° hopper and 210 mm for the 15° hopper.

The Mohr-Coulomb properties used in the FEM simulations should, preferably, be calibrated from experimental characterization. Unfortunately, these values were not reported in the Ketterhagen et al. work [44] and, therefore, these material properties are determined here from the DEM particle properties used by Ketterhagen et al. [44], which showed good quantitative agreement with experimental results. Specifically, the internal friction angle and wall friction angle were calibrated using DEM simulations of an annular shear cell (see [42] for details of this calibration procedure). Note that in the experiments by Ketterhagen et al. [44], the mass fractions of small particles were relatively small (10%). Hence, the particle diameter used in the DEM calibration simulations is identical to the large particle diameter with $d = 2.24$ mm. Material density, elastic modulus, and Poisson's ratio are known to have little influence on the granular flow behavior, and are assumed here to be the same as those obtained for hard spheres [42]. Dilation of cohesionless granular materials is usually small [36–38] and, thus, is also assumed here to be equal to the one obtained for hard spheres [42]. Finally, particle shape and size effects are lumped together with bulk and transport material properties used in the model. Tables 3 and 4 show the DEM and FEM material parameters, respectively, used in the hopper discharge simulations.

Table 3. Parameters used by Ketterhagen and co-workers [44] in DEM simulations.

| Parameter | Value |
|---|---|
| Particle density (kg/m$^3$) | 2520 |
| Particle-particle coefficient of restitution (-) | 0.94 |
| Particle-wall coefficient of restitution (-) | 0.90 |
| Particle-particle friction coefficient (-) | 0.1 |
| Particle-wall friction coefficient (-) | 0.5 |
| Rolling friction coefficient (-) | 0.045 |

Table 4. Parameters used in FEM hopper discharge simulations.

| Parameter | Value |
|---|---|
| Material density (kg/m$^3$) | 1500 |
| Young's modulus (MPa) | 3.65 |
| Poisson's ratio (-) | 0.065 |
| Internal friction angle (degree) | 18.4 |
| Cohesion (Pa) | 0 |
| Dilation angle (degree) | 0.1 |
| Wall friction coefficient (-) | 0.31 |

Since no velocity information was given in the previous work [44], the velocity profiles cannot be compared directly. However, there is extensive evidence that FEM models can accurately predict the velocity field of granular flows [36–38,42,43]. Therefore, the velocity profiles predicted by FEM simulations of the hopper geometries shown in Figure 2 are used to predict segregation during discharge.

As mentioned in Section 3, a single diffusion coefficient $D$, as opposed to an anisotropic tensor, leads to accurate predictions in segregation-dominated flows [21]. Also, as shown in Figure 7, segregation is dominated more by the percolation than the diffusion. Therefore, the diffusion coefficient calibrated by Liu et al. [43] in the spanwise direction is adopted and assumed homogeneous in the material domain. The value is determined by averaging the diffusion coefficient in the entire flowing layer of the bed after the flow becomes steady.

The percolation length scale $S$ is calibrated to one set of experimental data and used to predict other experimental configurations. Specifically, the aim is to reproduce the experimentally-observed normalized mass fraction of fines $x_i/x_f$, where $x_i$ is the mass fraction of fines collected at discharge in a given sample and $x_f$ is the initial mass fraction of fines in the bed. Figure 8 shows the normalized mass fraction of fines $x_i/x_f$ as a function of the fractional mass discharged $M/M_{total}$, where $M$ is the cumulative mass discharged and $M_{total}$ is the initial total mass inside the hopper. The experiment was performed in the 55° hopper with a well-mixed initial fill. The initial mass fraction of fines is 10% with the particle diameters for small and large particles equal to 1.16 mm and 2.24 mm, respectively. The multi-scale model predictions are also shown in Figure 8 with different assumed values of the percolation length scale $S$. Note that in the FEM simulations, due to the penalty contact algorithm used, as described in Section 2.2, the material tends to attach to the wall when almost fully discharged. Thus, the simulations are not a good description of the final stage of the discharge process. Regardless of this limitation, it is evident from the figure that the FEM-ADS model can predict the segregation pattern during hopper discharge with good qualitative accuracy.

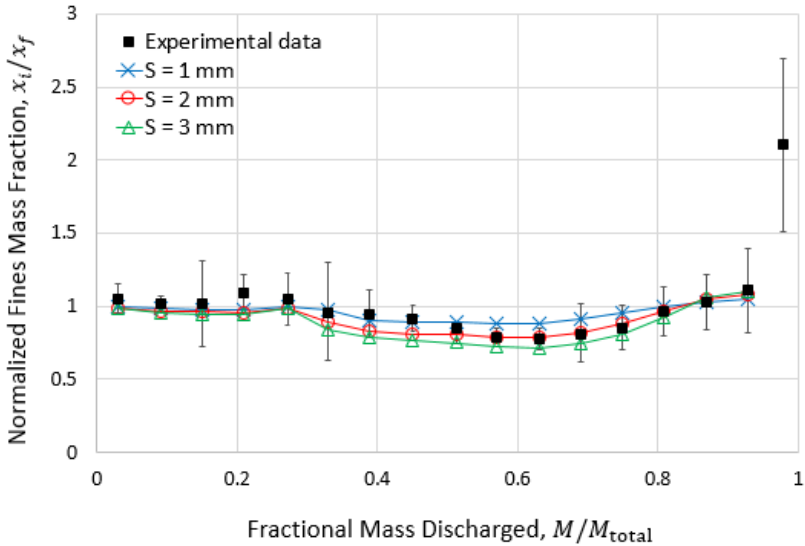

Figure 8. Experimental and FEM-ADS model predictions of the normalized mass fraction of fines with respect to the fractional mass discharged for different percolation length scale ($S$) values. The hopper angle is 55° and the initial fines mass fraction is 10%. Scatter bars denote the 95% confidence interval of the experimental results.

Figure 9 shows the calibration error for different values of $S$. The calibration error is defined as the averaged absolute differences in the normalized mass fraction of fines compared to the experimental results. The figure suggests that a value of $S = 2$ mm is optimal for the percolation length scale of the tested system. It is worth noting that this $S$ value is about seven times larger than the one used in Section 4.1 for the material system in the rotating drum. The reason for the difference may be that a smaller particle-particle friction coefficient is used in the hopper discharge simulation as compared to the rotating drum. A smaller particle-particle friction coefficient would make it easier for small particles to percolate through the large particles and, thus, give a larger percolation length scale.

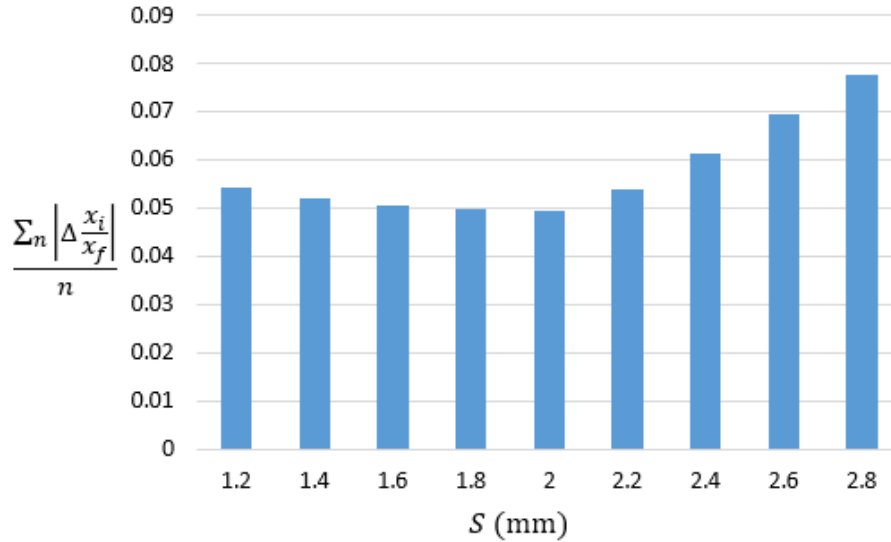

Figure 9. Averaged absolute differences between experimental values and FEM-ADS model predictions of the normalized mass fraction of fines, for different $S$ values.

The FEM-ADS model calibrated with data from the 55° hopper ($S = 2$ mm) is now compared to the experimental data from the 15° hopper using a well-mixed initial fill and 10% initial mass fraction of fines. The diffusion coefficient $D$ is equal to 0.6 mm$^2$/s for the 15° hopper discharge simulation. As mentioned previously, this value is determined by averaging the diffusion coefficient in the entire flowing layer in the 15° hopper. This value is different from the diffusion coefficient calibrated for 55° hopper since the velocity field changes between these two hoppers. Figure 10 depicts the small particle concentration spatial and temporal evolution. It is evident from the figure that segregation mainly occurs near the free surface where large particles tend to roll down the incline towards the hopper centerline. Since velocities are the largest near the centerline, these large particles are discharged first. This trend is in agreement with the experimental observations reported by Ketterhagen et al. [44] indicating that the multi-scale model qualitatively predicts the segregation pattern during hopper discharge.

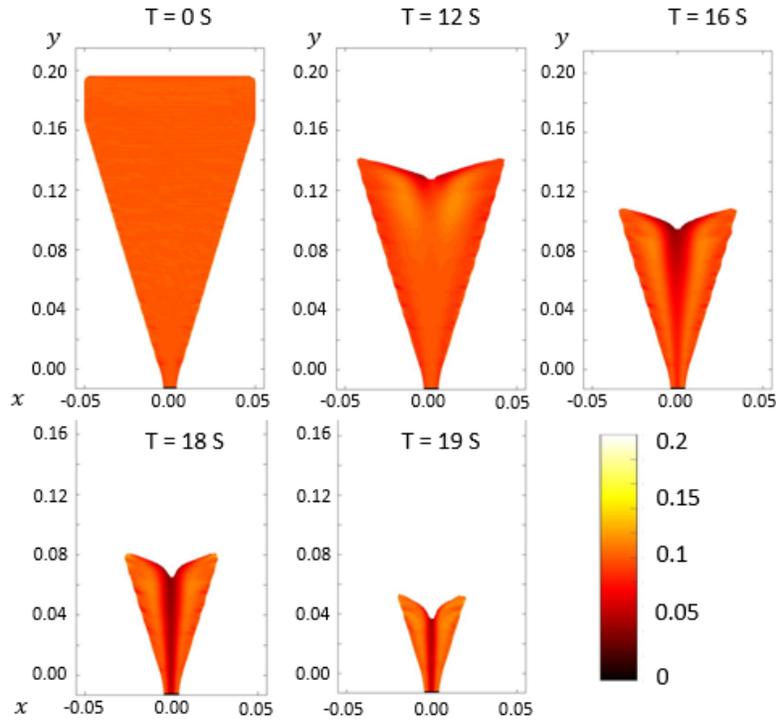
Figure 10. Simulation snapshots showing segregation evolution. The vertical color scale corresponds to the concentration of small particles.

Figure 11 shows the normalized mass fraction of fines as a function of the fractional mass discharged for the 15° hopper discharge experiment. As indicated previously, the ending stage of the discharge process is not shown in the figure because the FEM-ADS model is not appropriate for describing this stage. The figure shows good quantitative agreement between the model predictions and experimental measurements, at least within the experimental scatter. Moreover, different segregation patterns are observed in Figures 8 and 11 due to different flow modes. The 55° hopper primarily discharges in funnel flow while the 15° hopper primarily discharges with mass flow behavior. The reasons for these two different segregation patterns were discussed by Ketterhagen et al. [44].

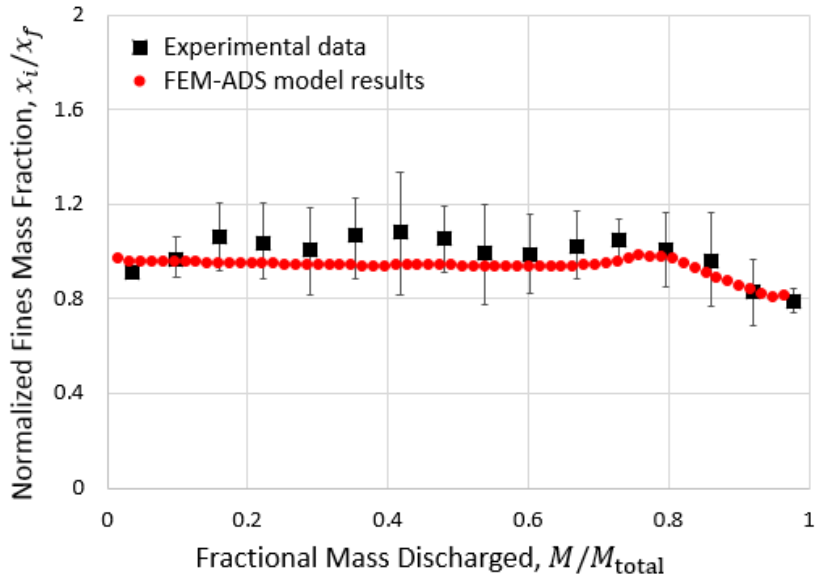

Figure 11. Experimental and FEM-ADS model predictions of the normalized mass fraction of fines with respect to the fractional mass discharged. The hopper angle is 15° and the initial fines mass fraction is 10%. Scatter bars denote the 95% confidence interval of the experimental results.

Note that to ensure the convergence of the ADS equation for the hopper simulations, a mesh dependency study was performed using the 15° hopper simulation. The normalized fine mass fraction, as shown in Figure 11, was computed using 250,000 and 640,000 elements, respectively. The averaged error among all computed data points is 0.36% and, hence, 250,000 elements were used in the hopper simulation studies.

Finally, the FEM-ADS model is further compared to experiments performed in the 55° hopper with a well-mixed initial fill, but different initial fines mass fractions, namely 20% and 50%. The same large and small particles are used in these experiments. The same diffusion coefficient ($D = 2.5$ mm$^2$/s for 55° hopper) and percolation length scale ($S = 2$ mm) as used previously are used in these new simulations. Figure 12 summarizes the good quantitative agreement between the experimental data and model predictions. The total wall-clock time required to run each of these simulations, including the FEM (32 cores with the Intel Xeon CPU E5-2680 v3 processor) and ADS (MATLAB, single core with the same processor) calculations, was two to three days.

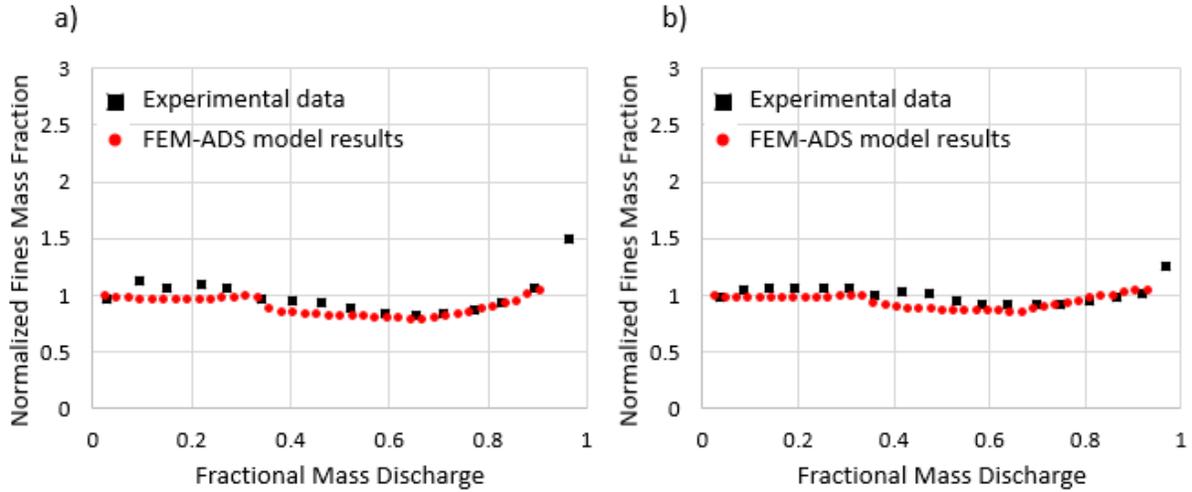

Figure 12. Experimental and FEM-ADS model predictions of the normalized mass fraction of fines with respect to the fractional mass discharged. The hopper angle is 55° and the initial fines mass fraction is (a) 20%, and (b) 50%.

## 5    Conclusions

A two-dimensional, transient modeling approach for predicting binary segregation in particulate systems was presented. This model is an extension of previously published works [42,43] and combines predictions from a finite element method (FEM)/Mohr-Coulomb elasto-plastic (MCEP) material model for the macroscopic granular velocity field with particle diffusion and segregation correlation parameters from calibration experiments or DEM simulations. The bulk velocity field and diffusion and segregation relations are combined in the model using the advection-diffusion-segregation (ADS) equation. The modeling approach was compared against segregation data from previously published rotating drum DEM simulations and discharging hopper experiments [22,44]. Since the MCEP material parameters were not reported in the rotating drum and hopper publications, these parameters had to be calibrated or estimated using DEM simulations of shear cell and uniaxial compression tests. In addition, the segregation length scale had to be calibrated for the hopper simulations from one of the hopper experimental data sets.

The FEM-ADS model segregation predictions were quantitatively accurate for both systems. A significant advantage of this multi-scale modeling approach is that it is expected to be more computationally efficient than DEM-only models for industrially-relevant system sizes [42,43]. Furthermore, all of the parameters used in the model can be measured from independent, standard tests or calibrated from simple two-dimensional experiments.

We close by pointing out future research directions and extensions of the multi-scale modeling approach. First, the current model is one-way coupled, which means that the material concentration is computed after the FEM simulation is completed. A two-way coupled model should be developed if materials with significantly different properties are used. Second, future work should focus on expanding the current multi-scale segregation model to three-dimensions [43] and more than two particle species, which are important in industrial practice. Lastly, standard experimental methods should be developed to calibrate the diffusion and segregation correlation parameters.